\documentclass[pdflatex, sn-mathphys]{SNmult}

\usepackage{type1cm}        
\usepackage{makeidx}         
\usepackage{graphicx}        
\usepackage{multicol}        
\usepackage[bottom]{footmisc}
\usepackage{tabularx}
\usepackage{longtable}
\usepackage{lscape}
\usepackage{amsthm}%
\usepackage{amsmath,amssymb,amsfonts}
\usepackage{cite}%
\usepackage{url}%
\usepackage{csquotes}
\usepackage{newtxtext}       %
\usepackage[varvw]{newtxmath}       
\usepackage{colortbl}

\makeindex             

\begin{document}
\title*{AI-Based Sound Effect Generation: A Narrative Review of Generative Models Across Input Modalities}
\titlerunning{AI-Based Sound Effect Generation} 
\author{Sandy Abdo\orcidID{0009-0006-5128-1625}, Bill Kapralos \orcidID{0000-0003-0434-3847}, Priyamvada Tripathi \orcidID{0009-0005-5070-7420}, KC Collins \orcidID{0000-0003-0695-7228}, and Adam Dubrowski\orcidID{0000-0002-2074-0933}}
\authorrunning{Abdo \textit{et al.}} 
\institute{Sandy Abdo \at Ontario Tech University, Oshawa, ON, Canada, \email{sandy.abdo@ontariotechu.net} \and Bill Kapralos \at Ontario Tech University, Oshawa, ON, Canada, \email{Bill.Kapralos@ontariotechu.ca}
\and Priyamvada Tripathi \at Durham College, Oshawa, ON, Canada, \email{Priyamvada.Tripathi@durhamcollege.ca}
\and KC Collins \at Carleton University, Ottawa, ON, Canada \email{kccollins@cunet.carleton.ca}
\and Adam Dubrowski \at Ontario Tech University, Oshawa, ON, Canada, \email{adam.dubrowski@ontariotechu.net}}
%
%
\maketitle
\abstract{Sound effects play a crucial role in conveying actions, events, and environmental cues across digital applications, often requiring a high degree of variation and contextual adaptability. Artificial intelligence (AI)-driven audio generative models are rapidly growing in popularity and have the potential to transform the way sound is synthesized and used across various applications. In response to this growing momentum, this chapter reviews and analyzes recent AI-based generative models for sound effect synthesis, with a focus on how different input modalities (text, visual, audio, and multimodal) affect the quality, controllability, and contextual relevance of the generated audio. It examines 30 peer-reviewed articles sourced from Google Scholar, IEEE Xplore, and the ACM Digital Library, exploring the evolution of AI generative models over the past five years. The results show that multiple models achieved state-of-the-art performance, producing high-fidelity, semantically aligned, and increasingly temporally coherent sound effects across tasks. However, despite these advances, the review identifies persistent challenges, including limitations in temporal synchronization for complex multi-event scenarios, gaps between objective metrics and human perception, and trade-offs between controllability and generative diversity. Overall, the chapter highlights that AI-driven sound effect generation is progressing toward more adaptive, scalable, and context-aware systems, offering significant implications for future sound design workflows and interactive media applications.
}


\section{Introduction}
\label{sec:1}
Sound is fundamental to how humans perceive, interpret, and interact with digital systems, and recent advances in artificial intelligence (AI) are transforming how audio is created, processed, and experienced. Rather than functioning as a static or purely reactive element, modern audio systems increasingly leverage AI to become adaptive, generative, and context-aware. This enables sound, including dialogue, sound effects, ambient noise, and music, to dynamically respond in real-time to user input, environmental conditions, and system states \cite{collins_game_2008}.
\par Among these, sound effects play a particularly critical role in conveying events, actions, and spatial cues, making them essential in applications such as video games, film, virtual simulations, and user interfaces where immediate auditory feedback enhances realism and user engagement \cite{collins_game_2008, serafin_sonic_2018}.

\par Designing dynamic and responsive sound effects, however, remains a complex and resource-intensive task. Sound designers must work through extensive audio databases, extract specific sound clips, and meticulously mix and manipulate sounds to achieve the desired auditory experience \cite{collins_game_2008}. However, digital applications including games and films may require thousands of sound effects to create experiences that resemble the real-world, especially since human pattern recognition capabilities make repeated sounds, such as footsteps, quickly noticeable and detrimental to immersion if not sufficiently varied \cite{collins_game_2008}. This repetitive workflow, combined with the need for constant variation, can be tedious and limit creative exploration. Automating some of these tasks using AI could streamline the process, allowing designers to focus on experimenting with alternative soundscapes and enhancing the overall audio experience \cite{filimowicz_doing_2022}.
\par As AI workflows become more integrated into digital applications, sound effect generation is increasingly shifting toward procedural and adaptive approaches \cite{steinmetz_audio_2025}. This makes it more difficult for sound designers to anticipate which sounds will be needed and under what conditions, as user interactions or system-driven events can produce unexpected and novel combinations of actions. Rare or unanticipated interactions may occur that were not considered during the design process, creating a demand for flexible and responsive audio systems. AI could synthesize plausible interaction sounds (e.g., action, background, and ambient) based on learned models of real-world acoustics \cite{steinmetz_audio_2025}. 
\par In such systems, sound must adapt dynamically not only to interactions but also to continuously changing conditions. Acoustic properties such as reverberation, occlusion, and layering may need to update in real-time as scenes evolve \cite{menexopoulos_procedural_2023}. Additionally, contextual parameters, such as user state, system variables, or desired emotional tone, can influence how sound effects are generated and modified. For example, a character’s footsteps could vary depending on factors like movement intensity, physical load, or condition, resulting in more nuanced and responsive auditory feedback. At the same time, procedural systems can generate vast numbers of variations in actions, objects, and events, making it impractical to design unique sound effects for every possibility \cite{menexopoulos_procedural_2023}. AI offers a scalable solution by dynamically generating and adapting sounds to match an effectively unlimited range of scenarios.
\par Beyond large-scale productions, many digital applications are developed by smaller studios, independent creators, and researchers who may lack the resources or expertise required to produce high-quality sound effects. AI-driven tools can help bridge this gap by lowering the cost and technical barriers to audio production, enabling more accessible creation of rich and context-aware sound \cite{filimowicz_doing_2022}. This is particularly valuable in specialized or underrepresented domains, where tailored sound design is needed but traditional resources are limited, allowing a broader range of creators to develop high-quality auditory experiences. 

\par The purpose of this chapter is to provide an overview of audio generative AI, and to identify the available tools and models in the current literature that can generate audio from prompts such as text, video, images, or other audio. This review maps current knowledge, identifies trends, and guides future work in audio generative AI. 
\section{Research Strategy}
This chapter is a review that follows the structure of a narrative literature review, as outlined by Oxman \textit{et al.} \cite{oxman_users_1994}. A narrative review serves to examine the literature comprehensively, offering a broad summary of the field. This approach is especially beneficial for readers new to the subject, providing them with valuable insights \cite{grant_typology_2009}. 

When crafting this chapter, we focused primarily on journal and conference papers, rather than for instance patents that may describe proprietary tools. The search was carried out using the Google Scholar, IEEE Xplore, and the ACM Digital Library databases between 6 February 2024, and 4 March 2025. We limited scope to these papers published within the past five years due to the rapidly changing nature of the field. Earlier models were included only for context and were not part of the search pool. IEEE Xplore, the ACM Digital Library, and Google Scholar were selected as the primary databases for this chapter due to their comprehensive coverage of research in computer science, engineering, and AI, fields most relevant to generative audio models. IEEE Xplore and ACM are leading publishers of conference proceedings and journals in AI and signal processing, providing access to cutting-edge research and original studies. Google Scholar complements these sources by offering broad, interdisciplinary coverage. Searches in other databases, such as Scopus and Web of Science, largely returned results already indexed in IEEE or ACM databases, indicating significant overlap and reinforcing the suitability of the chosen databases.

Additionally, the review focused on original, peer-reviewed research articles appearing in refereed journals and conference proceedings, written in English and present models capable of generating sound effects. Models that primarily generate speech or music were excluded, as these areas have already been extensively studied and are covered by several recent review articles \cite{liu_research_2023, kaur_conventional_2023, chen_musicldm_2024,dash_ai-based_2024, mitra_music_2025}. The research query was: (“Artificial intelligence” OR AI OR “language model” OR LLM OR “Deep Learning” OR “neural network”) AND (“Text-to-Audio” OR “video-to-audio” OR “visual-to-audio” OR “audio-to-audio” OR “image-to-audio”) AND (“sound effect” OR “audio effect” OR “game audio” OR “game sound”) AND (design* OR creat* OR generat* OR develop*). 
\subsection{Results}
The search initially yielded 204 articles. After removing duplicates and non–peer-reviewed articles \cite{sheppard_53_2020}, 60 abstracts were screened to assess relevance, specifically whether the study introduced an audio generative model. To ensure the relevance and quality of the reviewed literature, predefined inclusion and exclusion criteria were applied during both abstract screening and full-text review stages. Eligible studies were required to be peer-reviewed journal or conference papers published within the last five years, written in English, and presenting original research on artificial intelligence–based generative models capable of producing sound effects. Studies were excluded if they were non–peer-reviewed (e.g., preprints, theses, patents, or technical reports), review articles, surveys, or editorials. Additionally, works focusing primarily on speech or music generation, lacking a clear generative modeling component, or not producing audio as a primary output were omitted.

\begin{figure} [t]
    \centering
    \includegraphics[width=1.05\linewidth]{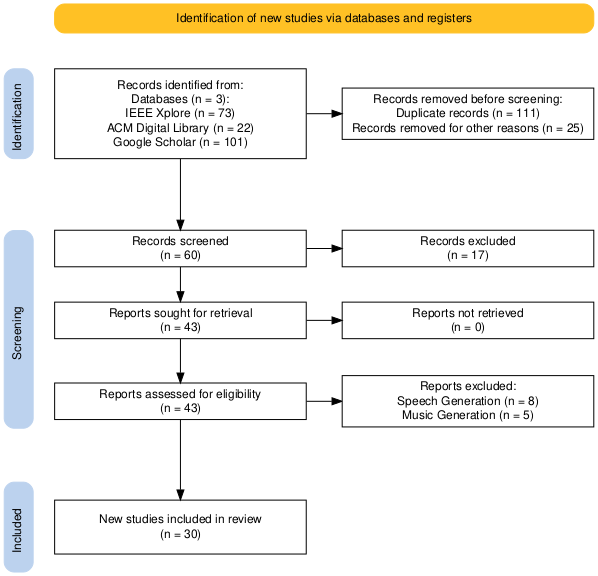}
    \caption{PRISMA flow diagram summarizing the review process.}
    \label{fig:prisma}
\end{figure}

Out of the 60 abstracts screened, 43 articles met the inclusion criteria. Following a detailed full-text examination, 30 of these were deemed relevant to the review, with further exclusions applied to studies lacking sufficient methodological detail or proper evaluation to ensure analytical rigor. Although this is a narrative review, the search and screening process was documented transparently, and a summary of the information flow through the phases of the review is provided in the Preferred Reporting Items for Systematic Reviews and Meta-Analyses (PRISMA) flow diagram \cite{haddaway_prisma2020_2022} shown in Figure \ref{fig:prisma}.

The remainder of this review is organized around four key themes identified in the literature: (i) text-to-audio models, which take text as input; (ii) visual-to-audio models, which take images or videos as input; (iii) audio-to-audio models, which use audio as input; and (iv) multimodal audio generative models, which incorporate multiple input modalities while producing audio as output. Each of these themes is discussed in a dedicated subsection. Table \ref{tab:Result_table} provides an overview of the models that will be discussed in this chapter.

\renewcommand{\arraystretch}{1.15}

\begin{longtable}{|>{\centering\arraybackslash}p{2.4cm}|
                  >{\centering\arraybackslash}p{2.4cm}|
                  >{\centering\arraybackslash}p{2.4cm}|
                  >{\centering\arraybackslash}p{5.2cm}|}

\caption{Summary of audio-generative models.} \label{tab:Result_table} \\

\hline
\textbf{Model} & \textbf{Architecture} & \textbf{Evaluation Metrics} & \textbf{Attributes} \\
\hline
\endfirsthead

\hline
\textbf{Model} & \textbf{Architecture} & \textbf{Evaluation Metrics} & \textbf{Attributes} \\
\hline
\endhead

\hline
\multicolumn{4}{r}{\textit{Continued on next page}} \\
\endfoot

\hline
\endlastfoot
\multicolumn{4}{|c|}{\textbf{Text-to-Audio}} \\ \hline
AudioLDM & LDM + CLAP & FD, IS, KL, OVL, REL & Efficient, scalable sound generation with high audio fidelity \\ \hline

Tango & LDM + Flan-T5 + VAE & FAD, KL, FD, OVL, REL & High audio quality with strong prompt relevance on small datasets \\ \hline

AudioLDM2 & LOA + LDM + AudioMAE & CLAPScore, FAD, KL, OVL, REL & Domain-agnostic model enabling semantic audio representation \\ \hline

SonifyAR & PbD+ LLM+ audioLDM & Preliminary usability evaluation & Interactive AR-based audio generation \\ \hline

Auffusion & Pixel VAE + LDM & FD, FAD, KL, IS, CLAPScore, OVL, REL & T2I method adaptation for TTA generation \\ \hline

Tango 2 & Diffusion-DPO + CLAP & FAD, KL, IS, CLAPScore, OVL, REL & Temporal precision and flexible audio generation across applications \\ \hline

SRC-gAudio & U-Net + LDM + VAE + HiFi-GAN & FD, KL, IS, FAD, CLAPScore, OGL, REL, AQ & Multi-sampling rate training for improved audio quality \\ \hline

Re-AudioLDM & Retrieval-Augmented LDM & IS, FAD, KL, CLAPScore & Retrieval-guided generation aligned with real-world sounds \\ \hline

Stable Audio Open & DiT +MLPs & FDopenl3 score, CLAPScore & Open-weight model for scalable audio generation \\ \hline

PicoAudio & U-Net + LDM & MOS, FAD, F1, $L_{1}^{freq}$ & Fine-grained temporal controllability \\ \hline

AudioComposer & LDM + hierarchical diffusion & F1, ACC, MAE, MOS & Hierarchical semantic modeling for improved audio quality \\ \hline

\multicolumn{4}{|c|}{\textbf{Video-to-Audio}} \\ \hline

V2RA-GAN & GAN & ODG, OSG, architectural evaluation, user study & Direct waveform synthesis via regression-based modeling \\ \hline

FoleyGAN & BigGAN + visual action recognition network & Retrieval Accuracy, IS, FID, NDB, user study & Visual-guided audio synthesis using spectrogram generation \\ \hline

FRIEREN & RFM + VAE + BigVGAN & FD, IS, KL, KID, FAD, Acc, MIS & Temporally aligned audio via ODE-based sampling \\ \hline

AutoSFX & SAM + spectrogram autoencoder + cross-attention & missing, redundant, mismatched sounds, rhythm alignment, acoustic similarity, user study & Pixel-level audiovisual feature-based sound generation \\ \hline

MIMOSA & Multi-step pipeline & Friedman, Wilcoxon, user study & Interactive spatial audio generation for content creators \\ \hline

FoleyGen & EnCodec + visual encoder + Transformer decoder & FAD, KLD, IB, OVR, REL, alignment & Cross-attention-based multimodal audio synthesis \\ \hline

SonicVisionLM & VLM + diffusion model & CLAP-top, Onset Accuracy, AP, Time Accuracy, IoU, IS, MKL, FID & Event-driven audio generation using vision-language modeling \\ \hline

MaskVAT & Transformer-based + RVQ codec & FDD, FDM, and FAD, semantic alignment, temporal synchronization & Masked token-based audio generation with temporal alignment \\ \hline

AV-LDM & MLP + VAE + LDM & FAD, AV-Sync, CLAPScore, user study & Ambient-aware controllable audio generation \\ \hline

Smooth-Foley & Auffusion + CLIP & FAD, MKL, CLIP Score, user study & Frame-level feature integration for improved alignment \\ \hline

STA-V2A & LDM TTA & FD, FAD, KL, IS, PAM, CLAPScore, AV, AA, OQ, AQ, SA, TA & Multi-level feature refinement for semantic and temporal alignment \\ \hline

LoVA & Autoregressive diffusion method & FAD, IS, MKL, audio quality, semantic relevance, consistency, overall evaluation & Long-form coherent and temporally consistent audio generation \\ \hline

TA-V2A & multi-modal approach + LLM + LDM & IS, FID, FAD, MKL, alignment, semantic, temporal alignment & Text-guided video-to-audio synthesis for improved coherence \\ \hline

\multicolumn{4}{|c|}{\textbf{Audio-to-Audio}} \\ \hline

NoiseBandNet & DDSP-derived filter-bank + MLP & MRSTFT, FAD, amplitude randomization, loudness transfer, user-defined control & Filter-bank-based synthesis for time-varying sound generation \\ \hline

CSTs (2 tools) & Mixed-Initiative Creative Interfaces + GAN & User Study (TA) & Interactive AI-assisted sound design for creative workflows \\ \hline

\multicolumn{4}{|c|}{\textbf{Multimodal}} \\ \hline

CoDi & Composable LDM + Latent alignment & FD, IS, REL & Unified latent-space multimodal generation, including audio generation \\ \hline

QueryMintAI & GPT-3.5 Turbo + DALL-E-2 + Whisper v2 + TTS-1 & TtPy, ASPy, HPy, ROUGE, user study & Integrated multimodal pipeline for versatile content generation \\ \hline

Amphion & AudioLDM + PicoAudio & FD, IS, KL & Open-source toolkit for text- and audio-driven generation \\ \hline

VAMG & GAN + Latent alignment & IS, FID, classification accuracy, user study & Bidirectional audio-visual generation with cross-modal alignment \\ \hline

\end{longtable}
\subsection{Evaluation Metrics for Audio Generation Models}
This section summarizes the objective and subjective metrics that will be presented in this chapter for evaluating AI-based sound effect generation models. Objective metrics provide quantitative measures of audio quality, diversity, and statistical similarity between generated and real-world data. Commonly used measures include distribution-based metrics such as Fréchet Distance (FD/FAD), Fréchet Inception Distance (FID), Kernel Inception Distance (KID), and KL Divergence (KL/KLD), as well as quality and diversity indicators including Inception Score (IS). Additionally, alignment and temporal accuracy are evaluated using metrics such as CLAP Score, F1 Score, and AV-Sync. In contrast, subjective metrics capture human perception of generated audio, focusing on aspects such as naturalness, fidelity, and alignment with input prompts. These include overall impression (OVL/OVR), mean opinion score (MOS), and audio quality (AQ/OQ), alongside alignment-focused measures such as audio-text relevance (REL), semantic alignment (SA), and temporal alignment (TA). Together, these metrics form a comprehensive evaluation framework that reflects both computational performance and perceptual quality (Table \ref{tab:Metric}).

\begin{table}[]
\caption{Evaluation metrics across models.}
\label{tab:Metric}
\begin{tabular}{lllllll}
\multicolumn{1}{c}{\cellcolor[HTML]{2E5D8C}{\color[HTML]{FFFFFF} \textbf{Objective Metric}}} & \multicolumn{1}{c}{\cellcolor[HTML]{2E5D8C}{\color[HTML]{FFFFFF} \textbf{Abbreviation}}} & \multicolumn{1}{c}{\cellcolor[HTML]{2E5D8C}{\color[HTML]{FFFFFF} \textbf{What It Measures}}} &  & \multicolumn{1}{c}{\cellcolor[HTML]{3A8A7C}{\color[HTML]{FFFFFF} \textbf{Subjective Metric}}} & \multicolumn{1}{c}{\cellcolor[HTML]{3A8A7C}{\color[HTML]{FFFFFF} \textbf{Abbreviation}}} & \multicolumn{1}{c}{\cellcolor[HTML]{3A8A7C}{\color[HTML]{FFFFFF} \textbf{What It Measures}}} \\
Fréchet Distance                                                                             & \textbf{FD / FAD}                                                                        & Audio   quality / realism                                                                    &  & Overall Impression                                                                            & \textbf{OVL / OVR}                                                                       & Perceived overall quality                                                                    \\
\cellcolor[HTML]{F0F2F5}Inception Score                                                      & \cellcolor[HTML]{F0F2F5}\textbf{IS}                                                      & \cellcolor[HTML]{F0F2F5}Diversity   + quality                                                &  & \cellcolor[HTML]{E8F5F2}Audio-Text Relevance                                                  & \cellcolor[HTML]{E8F5F2}\textbf{REL}                                                     & \cellcolor[HTML]{E8F5F2}Prompt alignment (human)                                             \\
KL Divergence                                                                                & \textbf{KL / KLD}                                                                        & Distribution   similarity                                                                    &  & Mean Opinion Score                                                                            & \textbf{MOS}                                                                             & Naturalness rating                                                                           \\
\cellcolor[HTML]{F0F2F5}CLAP Score                                                           & \cellcolor[HTML]{F0F2F5}\textbf{CLAPScore}                                               & \cellcolor[HTML]{F0F2F5}Text-audio alignment                                                 &  & \cellcolor[HTML]{E8F5F2}Audio Quality                                                         & \cellcolor[HTML]{E8F5F2}\textbf{AQ / OQ}                                                 & \cellcolor[HTML]{E8F5F2}Sound fidelity (human)                                               \\
F1 Score                                                                                     & \textbf{F1}                                                                              & Temporal event accuracy                                                                      &  & Semantic   Alignment                                                                          & \textbf{SA}                                                                              & Content   match (human)                                                                      \\
\cellcolor[HTML]{F0F2F5}Fréchet Inception Dist.                                              & \cellcolor[HTML]{F0F2F5}\textbf{FID}                                                     & \cellcolor[HTML]{F0F2F5}Feature-level quality                                                &  & \cellcolor[HTML]{E8F5F2}Temporal   Alignment                                                  & \cellcolor[HTML]{E8F5F2}\textbf{TA}                                                      & \cellcolor[HTML]{E8F5F2}Timing   accuracy (human)                                            \\
Kernel Inception Dist.                                                                       & \textbf{KID}                                                                             & Distributional fidelity                                                                      &  &                                                                                               &                                                                                          &                                                                                              \\
\cellcolor[HTML]{F0F2F5}AV-Sync                                                              & \cellcolor[HTML]{F0F2F5}\textbf{AV-Sync}                                                 & \cellcolor[HTML]{F0F2F5}Audio-visual synchronization                                         &  &                                                                                               &                                                                                          &                                                                                             
\end{tabular}
\end{table}
\section{Text-to-Audio Generation}
Text-to-Audio (TTA) is a growing field that recognizes, synthesizes, and generates audio based on given text input \cite{liu_research_2023}. To generate the audio, the provided text can consist of content, attribute, and style. The type of audio generated includes music, speech, and sound effects. The research in this area mainly focuses on text-to-speech (TTS), particularly the quality, efficiency, and control of speech sound \cite{liu_research_2023, kaur_conventional_2023}. Some progress has also been made in text-to-music (TTM) \cite{chen_musicldm_2024,dash_ai-based_2024, mitra_music_2025}. Unlike speech and music, sound effects are the least explored. In this chapter, we are focusing on research specifically examining sound effect generation, and therefore we do not describe speech or music generation here.

Liu et al. \cite{liu_audioldm_2023} introduced AudioLDM, a generation model that produces high-quality sound from text descriptions using a latent diffusion model (LDM) architecture. Unlike previous approaches that generate audio in the waveform space (a computationally intensive process), AudioLDM works in the latent space (a lower-dimensional representation of the data capturing essential features), enabling more efficient and scalable sound generation while preserving audio fidelity. The model leverages Contrastive Language-Audio Pretraining (CLAP), which provides a shared embedding space for both text and audio, allowing the model to understand and align textual prompts with audio features effectively.  Additionally, its training in the latent space of the audio coder greatly reduces computational cost without sacrificing perceptual quality. This also enables the model to generate audio that is longer and more diverse than previous methods. AudioLDM achieved state-of-the-art performance on several audio generation benchmarks, especially for text-to-audio generation tasks. The results show that AudioLDM models significantly outperform baseline TTA systems such as DiffSound, which can model high-dimensional audio signals, and AudioGen \cite{kreuk_audiogen_2023}, an autoregressive model trained on waveforms,  across both objective measures such as Fréchet distance (FD), inception score (IS) and Kullback–Leibler divergence (KL), and subjective evaluations, including overall impression (OVL), audio-text relevance (REL), with the best-performing variant, AudioLDM-L-Full, achieving the lowest FD and highest human-rated audio quality. These results indicate that AudioLDM generates audio that is more natural and better aligned with the textual prompts compared to baselines. The model is also capable of generating a wide variety of sounds, including environmental noises, music clips, and synthetic effects.

\par Ghosal et al. \cite{ghosal_text--audio_2023} introduced Tango, a LDM that leverages a Large language model (LLM) to enhance representational capabilities, fine-tuning, and the learning of complex concepts. Tango uses Flan-T5, an LLM, to generate mel-spectrogram tokens that represent the frequency content of an audio signal over time. These tokens are then processed by a pre-trained audio variational autoencoder (VAE) to reconstruct the mel-spectrogram and ultimately produce audio via a vocoder. Trained on the AudioCaps dataset \cite{kim_audiocaps_2019}, consisting of 45,438 audio clips with their paired caption, Tango outperformed previous state-of-the-art text-to-audio models including AudioLDM across both objective (Fréchet Audio Distance (FAD), KL, FD) and subjective (OVL, REL) metrics, through evaluations by six participants who rated the audio quality and its relevance to the input text. Although trained on smaller datasets, the model achieved particularly strong performance due to its use of the Flan-T5 text encoder, demonstrating better audio quality, relevance to prompts, and sample efficiency compared to baselines.
\par Liu et al. \cite{liu_audioldm_2024} introduced AudioLDM 2, a TTA generation tool capable of producing audio, music, and intelligible speech without the domain-specific biases that often limit broader applications, especially in complex scenarios. To overcome this limitation, the authors propose the \textquote{language of audio} (LOA), a vector representation of an audio clip’s semantic information. Unlike existing methods, LOA can theoretically capture both fine-grained acoustic details and coarse-grained semantic content. In other words, it can represent both what is being said, and the identity of the sound produced. AudioLDM 2 leverages a GPT-2 language model to translate this information into an audio-masked autoencoder (AudioMAE), a decoder pre-trained on diverse audio content. The processed information is then synthesized into audio using a LDM built on AudioMAE features. To evaluate its performance, the authors compared AudioLDM 2 against other audio-generation systems in TTA, TTM, and TTS tasks. The model achieving top scores across objective (CLAPScore, FAD, KL) and subjective metrics (OVL and REL), and significantly outperformed previous state-of-the-art models, including Tango, on AudioCaps and MusicCaps in TTA and TTM tasks, and achieved comparable performance in TTS. 
\par Su et al. \cite{su_sonifyar_2024} introduced SonifyAR, a context-aware audio generation tool designed to enhance sound effects in augmented reality (AR). The system leverages Programming by Demonstration (PbD) to detect potential sound-producing interactions in AR. These interactions are then analyzed by sound acquisition models and a LLM to generate a text-based description that includes details about the user, virtual object, and real-world environment. Next, the sound acquisition models and LLM identify the material and surface type, which are then used to create a text prompt for AudioLDM \cite{liu_audioldm_2023} to generate contextually appropriate sound effects. To evaluate the system, the authors conducted a preliminary user study with eight participants, who provided feedback on SonifyAR’s usability and effectiveness. Overall, participants responded positively, with many expressing a willingness to use the tool and acknowledged its potential to enhance AR immersion. 

\par Xue et al. \cite{xue_auffusion_2024} introduce Auffusion, a TTA model that integrates LDM from text-to-image (T2I) tasks to enhance cross-modal alignment. Unlike previous models that overlooked fine-grained performance, the authors incorporate pixel VAE, a technique commonly used in T2I to improve the quality of AI-generated images. Auffusion uses a text prompt to generate a latent representation, which is then reconstructed into an image by the VAE decoder. This image is subsequently denormalized into a mel-spectrogram and synthesized into audio. Their results show that Auffusion outperforms previous state-of-the-art models on the objective measures including FD, FAD, KL, IS, and CLAPScore, and generalizes well to the Clotho test set, a benchmark derived from the Clotho audio captioning dataset \cite{drossos_clotho_2020}, despite being trained on a much smaller dataset. Additionally, subjective evaluations including OVL and REL show superior text-audio alignment.

\par Majumder et al. \cite{majumder_tango_2024} introduced Tango 2, a TTA generative model designed to synthesize high-quality sound using deep learning techniques. Using a diffusion model, the system is trained on extensive datasets to learn intricate patterns and structures of audio, enabling realistic and coherent sound generation. With features such as high-fidelity synthesis, scalability, and potential real-time processing, the model offers flexibility across various audio applications, including music composition, speech synthesis, and sound design. The model uses CLAP, a model that learns acoustics from natural language supervision, and temporal perturbation strategies, allowing it to enhance its ability to generate semantically accurate and temporally coherent audio. The results indicate that Tango 2 outperforms both its predecessor Tango and the baseline AudioLDM2 across a wide range of objective and subjective metrics (REL and OVL), showing notable improvements in audio quality (FAD, KL, IS), semantic alignment (CLAPScore), and temporal precision (Strongly Temporally-Aligned evaluation Metric (STEAM) \cite{steam_2024}), especially with complex multi-event prompts. These gains are largely attributed to direct preference optimization (DPO) fine-tuning and event/temporal data augmentation strategies in the Audio-alpaca dataset, a dataset of prompts with preferred and less accurate audio outputs for training better TTA models (https://huggingface.co/datasets/declare-lab/audio-alpaca), which prove critical in enhancing performance and model preference alignment. These findings highlight the breakthrough potential of this approach, making it a powerful tool for various domains requiring high-quality audio generation.

\par Li et al. \cite{li_src-gaudio_2024} proposed a multi-sampling rate model called SRC-gAudio that uses sampling rate and text prompt as a condition to enhance audio generation quality. They mapped time step and sampling rate into a one-dimensional embedding, and along with text embedding they provided this conditioning information to a U-Net model, a neural network model used for image segmentation \cite{navab_u-net_2015}. The LDM then guided by the conditioned information, restores the audio latent to mel-spectrogram and audio waveform using VAE and HiFi-Generative Adversarial Network (HiFi-GAN) to generate audio. The evaluation of SRC-gAudio used objective metrics, FD, KL, IS, FAD, and CLAPScore, showing that joint training across sampling rates, most notably at 32 kHz and 48 kHz, improves performance compared to training separate models. Subjective evaluation, based on ratings from eight participants using OGL, REL, and audio quality (AQ), further confirms that pre-training on low-sampling-rate data boosts audio quality and that SRC-gAudio outperforms baselines such as AudioLDM2 in high-sampling-rate generation.

\par Yuan et al. \cite{yuan_retrieval-augmented_2024} present the Re-AudioLDM model designed to enhance audio generation by leveraging retrieved audio samples as references. It operates by first retrieving relevant audio samples from a large-scale text-audio dataset based on the input text. These retrieved examples serve as references that guide the synthesis process, ensuring the generated audio aligns more closely with real-world sounds. The model is comprised of two main components: i) a retrieval module that identifies similar audio samples using semantic similarity techniques, and ii) a generative model that synthesizes new audio based on the condition of the retrieved references. Experimental results demonstrate that Re-AudioLDM outperforms state-of-the-art models including AudioGen, AudioLDM, and Tango on all evaluation metrics (IS, FAD, KL, and CLAP score) when enhanced with retrieval information, especially excelling in generating semantically relevant and high-quality audio. It demonstrates strong performance on long-tailed and zero-shot audio generation tasks, outperforming traditional mixup strategies by effectively leveraging retrieved audio-text pairs to improve robustness and generalization. 

\par Evans et al. \cite{evans_stable_2025} created a publicly available open-weight TTA called Stable Audio Open that was trained on Creative Commons (CC) licensed audio. The LDM is a diffusion-transformer (DiT) \cite{evans_long-form_2024}, consisting of stacked blocks connected in series to attention layers and gated multi-layer perceptrons (MLPs), a type of neural network that consists of multiple interconnected nodes. To ensure no copyrighted content, the authors analysed the soundtracks before using them in their model training process. Results indicate that Stable Audio Open outperforms comparable baselines on the AudioCaps Dataset, achieving the best FDopenl3 score for AudioGen and AudioLDM2, indicating more realistic and plausible sound generation, while also scoring highest in CLAP score, reflecting strong alignment with text prompts. 

\par Xie et al. \cite{xie_picoaudio_2025} introduced PicoAudio, a diffusion-based TTA model capable of precise temporal controllability in audio generation. The model employs a latent diffusion process, similar to text-to-image models, but adapted for generating high-fidelity audio with fine-grained alignment to text prompts. It leverages a U-Net architecture conditioned on textual embeddings to iteratively refine audio samples, ensuring both semantic accuracy and precise timing. The temporal conditioning mechanism allows it to generate audio segments that align exactly with specified time intervals. This approach improves upon prior TTA models, which often struggle to maintain synchronization between textual descriptions and generated sound. Particularly, the models was tested using subjective metrics, including mean opinion score (MOS) assesses audio quality (naturalness, distortion, event accuracy) and temporal controllability (timestamp/frequency accuracy), with 10 evaluators rating five clips per model, and objective metrics (FAD for audio quality, segment F1 score for timestamp control, and a frequency error metric $L_{1}^{freq}$ for frequency control). The results show that PicoAudio significantly outperforms baselines such as AudioLDM2 and Audit, achieving precise temporal control and superior audio quality across single and multi-event tasks.
\par Wang et al. \cite{wang_audiocomposer_2025} introduced AudioComposer, a diffusion-based model designed for fine-grained audio generation using natural language descriptions. It integrates a transformer-based text encoder to process detailed semantic representations of sound attributes, such as pitch, rhythm, and texture, which guide the synthesis process. The model leverages a hierarchical approach, to generate audio using a fine-grained natural language description, allowing it to capture both high-level structures and intricate temporal details. It builds upon diffusion-based architectures, incorporating transformer-based text encoders to understand detailed descriptions, refine noisy signals and generate corresponding audio. AudioComposer was trained on a large-scale dataset with numerous audio samples, allowing it to generalize various sound types, including speech, music, and environmental noises. It significantly outperforms prior models in audio generation tasks involving timestamp, pitch, and energy control, across both objective metrics ($F_1$-score was used for event accuracy (ACC) for pitch and energy categorization, and mean absolute error (MAE) between pitch and energy was also evaluated), and subjective (MOS) metrics. This shows that the generated audio exhibits greater coherence, fidelity, and adherence to textual descriptions. Ablation studies confirm that its performance benefits come from its design, including flow-based diffusion and the use of diverse training data. The combination of hierarchical modeling, a powerful text encoder, and diffusion-based synthesis results in superior fidelity, better alignment with textual descriptions, and more flexibility in generating complex soundscapes. 

\par
\begin{figure}
    \centering
    \includegraphics[width=1.05\linewidth]{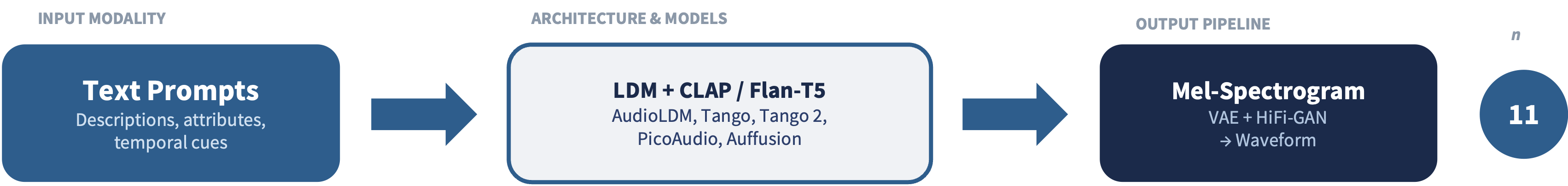}
    \caption{Dominant architectures and synthesis pipelines for TTA generation (n represents the number of studies/works in this group).}
    \label{fig:TTA}
\end{figure}

 In summary, recent advances in TTA generation, particularly for sound effects, demonstrate a rapid shift toward more efficient, controllable, and semantically aligned models (Figure \ref{fig:TTA}). Diffusion-based approaches operating in latent spaces, such as AudioLDM and its successors, have significantly improved audio fidelity and scalability, while the integration of large language models and retrieval mechanisms has enhanced the understanding of complex textual prompts and rare sound events. Emerging techniques further address key limitations by introducing temporal precision, multi-sampling strategies, and cross-modal learning, enabling more fine-grained and context-aware sound synthesis. Despite these advancements, sound effect generation remains less mature than speech and music synthesis, with ongoing challenges in realism, temporal alignment, and data efficiency. Overall, the reviewed work highlights both the substantial progress made and the promising future directions for developing robust, high-quality, and versatile sound effect generation systems.

\section{Visual-to-Audio Generation}
This section includes studies that use AI to generate audio from visuals such as images and videos. Video-to-audio (V2A) uses AI to recognize the content and style of a video to generate background music, spatial audio and sound effects \cite{liu_research_2023}. 
\par Liu et al. \cite{liu_towards_2022} present an end-to-end deep learning approach for generating raw audio from visual content using a Generative Adversarial Network (GAN)-based framework called V2RA-GAN. Unlike conventional methods that rely on spectrograms or physics-based models, this approach formulates sound synthesis as a regression problem, directly predicting synchronized raw audio from silent videos. The model includes a video encoder that extracts visual features, which are mapped to audio waveforms using a GAN. The system is fully trainable without additional inputs, improving scalability and reusability. The model was evaluated quantitatively using the Objective Difference Grade (ODG) for audio quality, and the Objective Similarity Grade (OSG) for audio similarity. Architectural evaluations indicated that filter layers and a combination of L1 and least-squares (LS) loss functions improved performance, while skip connections and Gaussian noise offered no benefits. Optimization schemes such as de-noising and peak-match improved clarity in specific scenarios, and user studies showed that V2RA-generated audio is generally high-quality and well-synchronized with visuals, often indistinguishable from real recordings. Overall, the results of the experiments demonstrated that V2RA-GAN produced high-quality, synchronized sounds in real-time, with practical applications in sound design and dubbing.
\par Ghose \& Prevost \cite{ghose_foleygan_2023} enhanced Foley generation by introducing FoleyGAN, a visually guided, class-conditioned model that uses a GAN designed to generate synchronous sound for silent videos. The model integrates a video action recognition network, which extracts temporal action information from video frames, and a sound generation network based on the BigGAN architecture. This architecture synthesizes high-resolution spectrograms conditioned on visual cues, which are then converted into audio using an inverse short-time Fourier transform (ISTFT). FoleyGAN was evaluated through qualitative and quantitative experiments. Quantitative metrics, including Sound Retrieval Accuracy, IS, Fréchet Inception Distance (FID), and Number of Statistically-Different Bins (NDB), show FoleyGAN with visual guidance outperforms baseline models in generation quality. Extensive ablation studies confirmed the effectiveness of temporal action information and the BigGAN architecture. Additionally, phase coherence analysis and a human survey demonstrated that generated sounds were generally perceived as high-quality and well-synchronized with visuals, with some variation across event classes. The model outperformed baseline models, achieving a sound retrieval accuracy of 76\%, high audio-visual synchronicity (81\% based on human surveys), and overall improved performance in comparison to other audio-visual synthesis methods.

\par Wang et al. \cite{wang_frieren_2024} introduced FRIEREN, a V2A model that uses rectified flow matching (RFM) to produce temporally aligned audio from silent video frames. It uses VAE to compress the mel-spectrogram in a continuous flow via vector field estimator, which is then converted to waveforms by the BigVGAN vocoder. RFM constructs straight-line flow trajectories, which allow for fast and accurate sampling when solved with an ordinary differential equation (ODE) solver. To allow for an audio-visual alignment, the vector field estimator uses a feed-forward transformer without temporal down-sampling, preserving the resolution of the temporal dimension. The model was evaluated quantitatively (FD, IS, KL, kernel inception distance (KID), FAD), and subjectively using MOS and outperformed prior models including Diff-Foley in inception score, audio fidelity, and alignment ACC, while being over nine times faster in one-step generation.

\par Wang et al. \cite{wang_autosfx_2024} introduced AutoSFX, a tool designed to automate sound design for videos. AutoSFX consists of two main components: i) sound generation, ii) and sound optimization. For sound generation, the tool follows the Segment Anything Model (SAM) \cite{kirillov_segment_2023}, extracting compact, pixel-wise audiovisual features from the video. A spectrogram autoencoder then predicts and generates the corresponding audio based on the provided visual prompt. For sound optimization, AutoSFX employs a cross-attention mechanism to align auditory and visual information, ensuring better synchronization between generated sounds and video content. To evaluate AutoSFX’s performance, the authors conducted a quantitative study using metrics such as missing, redundant, and mismatched sounds, rhythm alignment, and acoustic similarity, and compared the results to state-of-the-art methods. Results showed that AutoSFX achieved the lowest error rate among the tested models on the VEGAS \cite{zhou_visual_2018} and VGGSound datasets \cite{chen_vggsound_2020}. Additionally, AutoSFX outperformed the Diff-Foley model in generating animal sounds but was outperformed by Diff-Foley in generating instrumental sounds. The authors also conducted a user study with 30 participants, 10 video creators and 20 video viewers, who rated how well the generated soundtracks matched the visual content on a five-point scale. AutoSFX performed better than Pika, a free online audio generation tool, but fell short compared to professionally designed soundtracks. Despite this, 75\% of participants perceived the generated sounds as realistic, 80\% of viewers considered AutoSFX \textquote{very helpful} for amateur content creators, and 30\% of video creators found it \textquote{very helpful} for automating complex sound design processes.

\par Ning et al. \cite{ning_mimosa_2024} created an audio generation tool called Magnifying Immersion by Manipulating Objects in Spatial Audio (MIMOSA) to help amateur content creators generate spatial sounds (categorized as 3D localization or the creation of immersive experience) for videos. Their model uses a multi-step pipeline consisting of object detection, depth estimation, sound separation, audio tagging, and spatial audio rendering. This allows the users to interact with the tool, validate and correct errors in the AI-generated spatial audio by adjusting visual overlays in 2D and 3D manipulation panels. The model was evaluated by eight evaluators who rated five types of audios (Raw, Monaural, Default MIMOSA, and User-edited MIMOSA) on immersion and realism Statistical analysis using the Friedman test and Wilcoxon signed-rank tests. Results revealed that User-edited MIMOSA achieved significantly higher immersion than all other types, while default MIMOSA achieved immersion ratings comparable to raw audio. For realism, raw audio was rated highest, with computationally generated audio showing reduced realism due to perceived distortion. A user study was also conducted to examine how the tool assisted 15 novice content creators during their audio editing process. MIMOSA received high scores across metrics: usefulness, immersiveness, expressiveness, and ease of use. Participants praised the system's real-time feedback, intuitive 2D/3D manipulation panels, and visual cues for error detection. While most favored graphical manipulation, some preferred numerical inputs, indicating flexible user needs. Suggestions included better object selection tools and an action history panel for improved editing workflow. Participants found the tool easy to use, supports creativity, and improves immersion. They also concluded that allowing the users to refine the AI generated audio significantly enhances immersion while maintaining realism, highlighting the importance of human-AI collaboration.

\par Mei et al. \cite{mei_foleygen_2024} introduced FoleyGen, an advanced V2A generation model designed to produce high-quality, visually aligned soundscapes. Built on a language modeling framework, it incorporates EnCodec, a neural audio codec for bidirectional waveform-token conversion, a visual encoder for feature extraction, and a Transformer-based audio decoder. Two variants, FoleyGen-C and FoleyGen-P, differ in their handling of visual features, FoleyGen-C employs a cross-attention mechanism, while FoleyGen-P integrates features via self-attention. FoleyGen was evaluated using both objective metrics, FAD, KLD, and ImageBind (IB) score, and subjective human evaluations assessing overall quality (OVR), REL, and alignment. FoleyGen outperformed prior models (SpecVQGAN and IM2WAV) across all metrics, achieving lower FAD and KLD scores and higher IB scores. Specifically, FoleyGen-C showed superior performance compared to FoleyGen-P, due to enhanced cross-attention mechanisms and better integration of visual cues. The best results were achieved using all-frame visual attention and multi-modal pretrained encoders such as CLIP and IB. Experiments with different visual encoders and attention mechanisms highlighted the advantages of multi-modal pretraining and all-frame attention in achieving better temporal alignment. Despite its success, challenges in perfect audio-video synchronization remain, suggesting directions for future improvements.

\par Xie et al. \cite{xie_sonic_2024} proposed a  novel framework, SonicVisionLM, designed to generate audio for silent videos using vision-language models (VLMs). Instead of directly synthesizing audio from visual data, the model first identifies relevant events in a video using a VLM, which then suggests appropriate sound effects. This approach simplifies the complex task of aligning video and audio by breaking it down into well-studied subproblems: image-to-text and text-to-audio mapping, using diffusion models. A key innovation is the introduction of a time-controlled audio adapter, ensuring better synchronization of generated sounds with video events. The SonicVisionLM model was evaluated using the Greatest Hits and CountixAV datasets across both conditional and unconditional sound generation tasks. For conditional generation tasks, it outperformed the prior state-of-the-art model on metrics such as CLAP-top, Onset Accuracy, Onset Average Precision (AP), Time Accuracy, and Intersection over Union (IoU). For unconditional generation, the model surpassed existing video-to-audio generation models in IS and Mean KL (MKL), especially on the Greatest Hits dataset, using the IS, FID, MKL, and IoU metrics. Subjective evaluations further confirmed its superior performance in audio quality, temporal alignment, and synchronization. 
\par Pascual et al. \cite{leonardis_masked_2025} propose masked generative video-to-audio transformers (MaskVAT), a model that generates audio from silent videos using a Transformer-based architecture that predicts tokenized audio sequences via masked generative token modeling. The model employs a Residual Vector Quantization (RVQ) codec, a hierarchical tokenizer to compress the audio and embedded using pre-trained Descript Audio Codec (DAC) and return a codegram. The authors tested three variants of the model: i) $MaskVAT_{AdaLN}$, which uses AdaLN blocks, a modulator used in diffusion transformers, that matches the length of the transformer input using interpolation; ii) $MaskVAT_{Seq2seq}$, a sequence-to-sequence model that uses a transformer encoder, bidirectional encoder representation from audio transformers (BEATs), for visual embeddings and then uses cross-attention blocks, a parallel decoder, to mix the conditions with the main token, allowing auxiliary loss for synchronization; and iii) $MaskVAT_{Hybrid}$, combining both strategies where BEATS encoder to process visual features and AdaLN to align derived semantic features and S3D-derived alignment-sensitive features. The model was evaluated both objectively and subjectively on audio quality (FDD, FDM, and FAD), semantic alignment (novelty score and SparseSync), and temporal synchronization. Objective evaluation demonstrated that MaskVAT models, particularly the Seq2Seq and Hybrid variants, excelled in full-band audio quality. For semantic relevance, Seq2Seq and Hybrid variances resulted in the highest CLIP-Based similarity scores whereas alignment metrics resulted in higher scores for variants with AdaLN. Subjective evaluation confirmed these trends and indicated that $MaskVAT_{Hybrid}$ was rated highest in alignment and overall quality, and competitive with V2A-Mapper in fidelity and relevance, particularly among expert listeners. Overall, the results indicated that $MaskVAT_{Hybrid}$ as the most balanced and effective model for generating temporally aligned, semantically coherent, high-quality audio from video.

\par Chen et al. \cite{leonardis_action2sound_2025} introduced AV-LDM, an ambient aware audio generative model that generates sounds from silent egocentric videos. The model introduces a training strategy that conditions audio generation on neighbouring audio clips to factor out ambient sounds. The audio waveform is first converted into a mel-spectrogram and then compressed into a latent representation using a VAE encoder. For conditioning, audio from a different timestamp (the audio condition) is processed using the same VAE encoder and further transformed into a fixed-length vector using a multilayer perceptron (MLP). Similarly, the input video is passed through a pre-trained video encoder to extract relevant visual features. These combined audio and video features are used to condition a latent diffusion model, enabling controllable generation of ambient sounds and enhancing the synthesis of action-focused audio. The model’s performance was evaluated both objectively, using FAD, Audio-visual synchronization (AV-Sync) and CLAP scores, and subjectively using human evaluation. Results showed that the model outperformed baselines such as Diff-Foley, Spec-VQGAN, REGNET, and retrieval-based models. Human evaluation with 20 participants further confirmed AV-LDM's superior synthesis of semantically relevant, synchronized action sounds with controllable ambient noise, demonstrating promising generalization to VR game environments.

\par Zhang et al. \cite{zhang_smooth-foley_2025} proposed Smooth-Foley, a V2A generation model with improved semantic (alignment of generated sound with video content) and temporal (audio that is synchronized with the video) alignments. The authors established this by improving the temporal condition’s accuracy by using textual labels and by using a high-resolution frame-wise video embeddings instead of clip-wise. They integrated Auffusion, a TTA model with two lightweight adapters: i) a frame adapter, ii) and a temporal adapter (Xue et al., 2024). The frame adapter improves semantic and temporal alignment by incorporating high-resolution frame-wise video features instead of clip-wise embeddings. The temporal adapter refines synchronization by computing similarities between video frames and textual labels using CLIP, providing more accurate temporal conditions. These adapters were trained separately while keeping the base TTA model frozen, allowing efficient adaptation. The model was evaluated used both objective and subjective evaluations. Objective measures included FAD, MKL, and CLIP Score. Subjectively, ten expert evaluators rated models on semantic alignment, temporal alignment, and audio quality. Smooth-Foley outperformed baseline models on both objective and subjective metrics across VGG and VGG-C datasets, with frame-wise features, yielding the best results in semantic alignment, audio fidelity, and temporal synchronization. Qualitatively, the model captured complex audio events more accurately and demonstrated superior handling of temporal dynamics and realism.

\par Ren et al. \cite{ren_sta-v2a_2025} addressed semantic and temporal alignment challenges with their STA-V2A model. They employed an LDM TTA framework and a cross-modal guidance approach that integrated both text and video to ensure proper alignment. To mitigate the interference of redundant information, they refined features at both local and global levels. For local refinement, they used a pre-text task that predicted pseudo-labels of audio onset from video, enabling the acquisition of localized video features. For global refinement, they extracted semantic features from videos using an attentive pooling module. The model's performance was evaluated using both objective and subjective measures. Objective evaluations used the FD, FAD, KL, IS tools, prompting audio-language models (PAM), CLAP, audio-video alignment (AV-Align), and audio-audio alignment (AA-Align). Results showed that STA-V2A achieves the lowest FAD and strong scores across the board, though it trails slightly behind FoleyCrafter, another V2A model, in PAM. Subjective evaluations by six expert annotators on quality (OQ), audio (AQ), semantic (SA), and temporal alignment (TA) confirmed that STA-V2A resulted in the highest scores and narrow 95\% confidence intervals. Ablation studies further demonstrate that integrating filtered data, ControlNet with video features, onset-driven local features, and global semantic cues significantly enhances performance in audio quality, semantic alignment, and especially temporal synchronization. Overall, the STA-V2A model outperformed all baseline models across most objective and subjective metrics, including Diff-Foley.

\par Cheng et al. \cite{cheng_lova_2025} present Long-form Video-to-Audio (LoVA), a model designed for long-form video-to-audio generation. It integrates a video encoder with an autoregressive audio generator to produce coherent and temporally aligned soundtracks for extended video sequences. The model is trained on diverse video-audio datasets, leveraging self-supervised learning techniques to enhance synchronization and realism. LoVA was tested using both objective metrics, FAD, IS, and MKL, and subjective human evaluations across four aspects (audio quality, semantic relevance, consistency, and overall). Results demonstrated that LoVA outperforms existing methods in generating high-quality, context-aware audio, maintaining consistency over long durations and requiring the fewest inferences per audio. Experiments indicated its effectiveness in producing naturalistic sounds that closely match video content, setting a new benchmark for long-form audio generation. Overall, LoVA achieved best overall performance, excelling in both quantitative and qualitative assessments.

\par You et al. \cite{you_ta-v2a_2025} designed TA-V2A, a textually assisted video-to-audio generation model, that creates realistic audio by leveraging both visual and textual inputs. It was developed using a multi-modal learning approach, integrating deep learning techniques to synthesize coherent soundscapes based on video content and supplementary textual descriptions. The model was evaluated using both objective and subjective assessments to benchmark its audio generation performance. Objective evaluation was conducted using four metrics, IS, FID, FAD, and MKL, on the VGGSound dataset, generating 8 s clips. Additionally, alignment accuracy was used to measure audio-video synchronization. TA-V2A, especially when using video-audio-language pretraining (CVALP) with concatenation, outperformed baseline models across all metrics, showing strong semantic fidelity and alignment, with particularly low FID and FAD values. In the subjective evaluation, 20 participants rated the audio quality using MOS for both semantic consistency and temporal alignment. TA-V2A with user-provided prompts achieved the highest scores, demonstrating that text-controlled inference significantly enhances perceived audio-video coherence. 

\par
\begin{figure}
    \centering
    \includegraphics[width=1.05\linewidth]{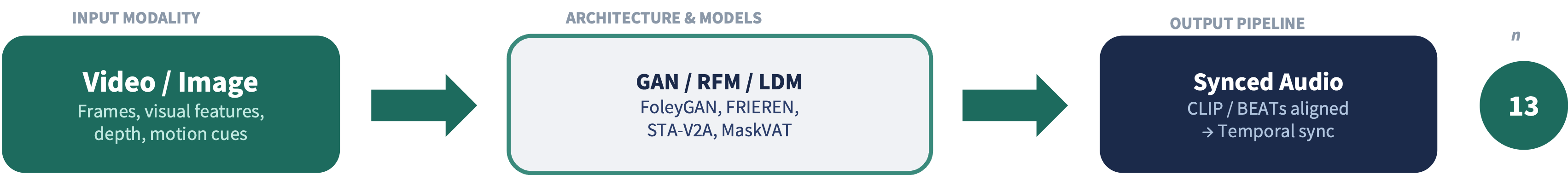}
    \caption{Dominant architectures and synthesis pipelines for V2A generation (n represents the number of studies/works in this group).}
    \label{fig:Video}
\end{figure}

 In summary, the reviewed studies demonstrate rapid progress in V2A generation, with modern approaches achieving increasingly high levels of audio quality, semantic relevance, and temporal synchronization (Figure \ref{fig:Video}). Techniques such as GANs, diffusion models, and transformer frameworks have each contributed unique strengths, from real-time waveform synthesis and high-fidelity spectrogram generation to improved alignment through cross-modal guidance and textual conditioning. Notably, recent work emphasizes not only automation but also controllability and human-AI collaboration, as seen in systems that incorporate user feedback or textual prompts to refine outputs. Despite these advances, challenges remain in achieving perfectly realistic soundscapes and flawless synchronization across diverse and complex scenarios. Overall, the literature highlights a clear trajectory toward more robust, scalable, and user-adaptive V2A systems, with promising applications in content creation, virtual environments, and automated sound design.

\section{Audio-to-Audio Generation}
Audio-to-audio generation in AI refers to systems that take an existing audio signal as input and produce a transformed or newly synthesized audio signal as output, rather than generating sound from text or images. It is often used to create variations of a reference sound or to apply stylistic or semantic changes while preserving key characteristics of the original audio \cite{liu_research_2023}.

\par Barahona-Ríos \& Collins \cite{barahona-rios_noisebandnet_2024} present NoiseBandNet, a novel neural audio synthesis method designed for generating time-varying sound effects. It builds on Differentiable Digital Signal Processing (DDSP) \cite{engel_ddsp_2020} (an architecture that allows the integration of a signal processing component with deep learning techniques) but replaces the harmonic-plus-noise synthesizer with a filter-bank-based approach. This architecture processes white noise through a precomputed multi-band finite impulse response (FIR) filterbank, generating M-band noise bands that span the entire frequency spectrum. The neural network predicts time-varying amplitude values for each band, conditioned on high-level audio features such as loudness and spectral centroid. During synthesis, these amplitudes are up-sampled and applied to the noise bands, which are then summed to reconstruct the target audio. This model is trained to predict time-varying amplitudes for multiple noise bands, which are then combined to reconstruct the target audio. The study evaluated NoiseBandNet's performance against various DDSP configurations using two objective metrics, i) Multiresolution Short-Time Fourier Transform (MRSTFT) loss, and ii) FAD, and found that NoiseBandNet significantly outperforms all four traditional DDSP noise synthesizers variants in MRSTFT loss. Although FAD followed similar trends, DDSP512 taps slightly outperformed NoiseBandNet on the pottery category, likely due to small dataset artifacts. Additional experiments, including amplitude randomization, loudness transfer and user-defined control parameters demonstrated creative sound design applications, highlighting NoiseBandNet's superior performance and flexibility for expressive, controlled sound synthesis.

\par Kamath et al. \cite{kamath_sound_2024} created two AI-based Creative Support Tools (CSTs), which are AI models that allow sound designers to interact with and experiment with the StyleGAN model to assist with their creative practice. Their goal was not to compare these two interfaces, but rather, to provide expert sound designers with two AI tools to interact with. The first interface allows sound designers to generate a more realistic sound by inputting a synthetic sound designed by manipulating acoustic parameters (e.g., frequency and impulse width). The second interface uses technology-specific controls to generate the edited sound \cite{weisz_toward_2023}. Using a qualitative approach, the authors conducted semi-structured interviews where they asked nine professional sound designers to provide their feedback on each of the CSTs. To analyze the interview transcripts, the authors used reflexive thematic analysis (TA), employing Atlas.TI (https://atlasti.com) for semantic coding and affinity diagramming to organize 76 codes into 12 themes. Rather than aiming for data saturation, the study prioritized participant diversity. The analysis revealed that the tool enables rapid iteration, generation of unique sounds, and minimizes reliance on manual recordings. While the tool’s unpredictability can spark creative ideas, it can also make it harder to complete tasks that require precision. Designers often prioritize artistic expression, even when the AI is focused on producing realistic sounds. As a result, although they value the creative possibilities the tool offers, they stress the importance of having more control and predictability, as well as tools that support, rather than replace, their creative decision-making.

\par
\begin{figure}
    \centering
    \includegraphics[width=1.05\linewidth]{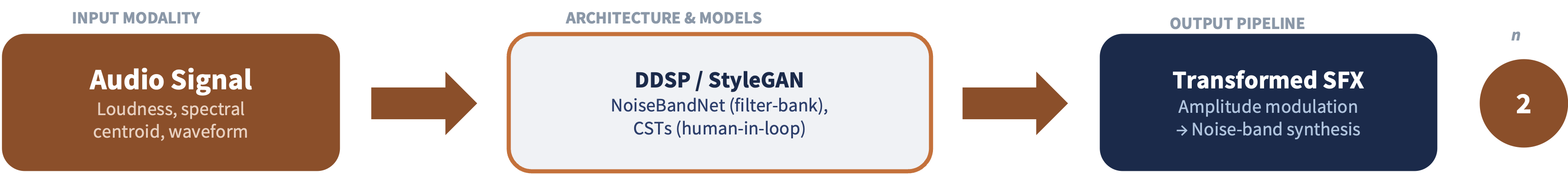}
    \caption{Dominant architectures and synthesis pipelines for audio-to-audio generation (n represents the number of studies/works in this group).}
    \label{fig:audio}
\end{figure}

 In summary, recent advancements in audio-to-audio generation demonstrate a clear shift toward more expressive, controllable, and application-oriented sound synthesis systems (Figure \ref{fig:audio}). Approaches such as NoiseBandNet highlight how integrating signal processing principles with deep learning can significantly improve synthesis quality and flexibility, particularly for complex, time-varying sounds. At the same time, the development of creative support tools underscores the importance of human-centered design, where AI serves as a collaborator rather than a replacement. Together, these works suggest that the future of audio-to-audio generation lies not only in improving technical performance, but also in enhancing user control, interpretability, and creative empowerment, enabling more seamless integration of AI into professional sound design workflows.

\section{Multimodal Audio Generation}
This section discusses approaches that employ various forms of user input to generate various outputs including audio. Since our goal is audio generation tools, we focus on aspects of the model that generate audio output.

\par Tang et al. \cite{tang_any--any_2023} introduced Composable Diffusion (CoDi), a multimodal generative model that can generate any combination of output modalities (text, image, audio, and video) from any combination of input modalities. The model employs a LDM that is trained independently and then integrated through the ‘Bridging Alignment’ technique that employs text as an anchor modality to align conditional encoders. Additionally, the modalities are synchronized using a technique called latent alignment that allows a modality-specific environment encoder into a shared latent space of other modalities. The model was tested objectively under single and multi-condition generation. For single-condition generation, each modality was tested separately. Audio generation uses a VAE decoder and a vocoder to reconstruct audio samples from mel-spectrograms. In multi-condition generation, where the model generates two modalities simultaneously, it employs a technique called latent alignment. This allows each modality-specific encoder to project into a shared latent space, enabling cross-modal interaction. Additionally, a cross-attention layer is added to each modality’s U-Net, allowing each to access the latent features of the others. The model was evaluated quantitatively using metrics such as FD, IS, and Relevance (REL), demonstrating superior fidelity, relevance, and cross-modal understanding.

\par Ghosh \& Deepa \cite{ghosh_querymintai_2024} introduced QueryMintAI, a multimodal LLM designed to process various types of user inputs (text, images, videos, documents, URLs, audio, and databases) and generate corresponding outputs across different modalities. It leverages OpenAI’s GPT-3.5 Turbo (text-to-text generator), DALL-E-2 (text-to-image generator), Whisper v2 (speech-to-text), and TTS-1 (text-to-speech generator), among others, to facilitate seamless interactions, allowing it to generate various forms of outputs. Through the model, the authors aimed to enhance user experience by providing a unified, intelligent, and private AI assistant capable of handling diverse data formats. They compared the model to other leading AI systems, including ChatGPT (GPT-3.5), Gemini, CoPilot, and ClaudeAI. The results showed that QueryMintAI outperformed these models in both multimodal capabilities and overall performance, as measured by key metrics such as Total Text Perplexity (TtPy), Average Sentence Perplexity (ASPy), Highest Perplexity (HPy), and ROUGE scores (i.e., ROUGE-1 (R1), ROUGE-2 (R2), and ROUGE-L (RL)). User evaluation with 50 participants assessed the relevance, fluency, coherence, and overall quality of the model, which were aligned with the objective results. Overall, the study concluded that QueryMintAI is a highly versatile, multimodal AI model that delivers improved performance across multiple domains, making it a powerful tool for personal AI applications, accessibility, and creative content generation.

\par Zhang et al. \cite{zhang_amphion_2024} introduced Amphion, a user-friendly, open-source toolkit designed for generating audio, speech, and music. Amphion aims to help both beginners and researchers explore generative AI with ease. It supports multiple tasks, including TTS, TTM, TTA, and audio-to-audio generation. We focus on TTA and audio-to-audio generation. For TTA, Amphion uses pre-trained models like AudioLDM and PicoAudio to generate audio from descriptive text. The model achieved better performance on FD, IS, and KL scores in comparison to open-source models such as Diffsound and AudioLDM. The model also supports audio-to-audio generation, referred to as wavelength-to-wavelength, processes auditory input to perform tasks such as voice conversion, singing voice conversion, emotion conversion, accent conversion, and speech translation. In their paper, the authors only evaluated singing voice conversion where they had the model convert 48 singing utterances into a female and a male target. Results showed that the model outperformed SoftVC (https://github.com/bshall/soft-vc)  in both naturalness and speaking similarity. To evaluate its performance, Amphion was compared against other open-source models and found that it consistently delivered superior results.

\par Hao et al. \cite{hao_vag_2025} present the visual-audio mutual generation (VAMG),  dynamic cross-modal model designed to compensate for missing audio or visual information in videos. The model facilitates both audio-to-visual and visual-to-audio generation while also allowing self-generation within each modality. It uses GANs with joint optimization of modal reconstruction and adversarial constraints to address issues of structural alignment and signal compensation. The trained model demonstrated effectiveness in instrument- and pose-oriented audio-visual generation. VAMG significantly outperformed prior methods for generating images from corresponding sounds (S2IC) and for generating sounds from corresponding images (I2S) in generation quality, as quantified by IS, FID, and classification accuracy. Additionally, subjective evaluation by 21 participants showed significant improvements in the satisfactory and acceptable ratings of the generated modality compared to baseline. Through latent embeddings and a novel adversarial loss function, VAMG enables dynamic modality compensation, ensuring more accurate and coherent cross-modal translations.

\par
\begin{figure}
    \centering
    \includegraphics[width=1.05\linewidth]{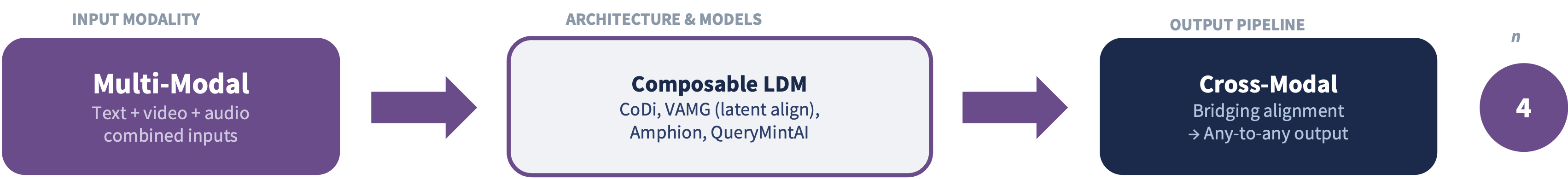}
    \caption{Dominant architectures and synthesis pipelines for multimodal generation (n represents the number of studies/works in this group).}
    \label{fig:multimodal}
\end{figure}

 In summary, these models highlight a clear trend toward increasingly flexible and unified multimodal systems that incorporate diverse user inputs to produce high-quality audio outputs (Figure \ref{fig:multimodal}). Models such as CoDi emphasize tightly coupled latent representations for synchronized cross-modal generation, while frameworks like QueryMintAI demonstrate the effectiveness of integrating specialized models into a cohesive pipeline for practical applications. Toolkits such as Amphion further lower the barrier to entry by providing accessible implementations of advanced audio generation techniques, and approaches like VAMG push the boundaries of cross-modal reasoning by enabling bidirectional generation and compensation between modalities. Collectively, these models illustrate that modern audio generation systems are moving beyond isolated tasks toward holistic, multimodal frameworks that leverage shared representations, cross-modal alignment, and user-centric design to achieve more coherent, context-aware, and versatile audio synthesis.

\section{Discussion}
This chapter highlights the rapid advancement of AI-driven generative models for sound effect synthesis, particularly over the past five years. Across text-to-audio, visual-to-audio, audio-to-audio, and multimodal systems, a clear trend emerges toward increasingly sophisticated models that can generate high-quality, contextually relevant, and temporally aligned audio (Table \ref{tab:summary}). Diffusion-based architectures, especially latent diffusion models (LDMs), consistently demonstrate strong performance across multiple modalities, largely due to their ability to balance computational efficiency with high-fidelity output \cite{liu_audioldm_2023, liu_audioldm_2024}.

One of the most significant findings is the improvement in semantic alignment between input prompts and generated audio. Models such as AudioLDM2, Tango 2, and PicoAudio demonstrate that integrating language models and multimodal embeddings enhances the system’s understanding of contextual cues, resulting in more accurate and expressive sound generation \cite{xie_picoaudio_2025, liu_audioldm_2024, majumder_tango_2024}. Similarly, in visual-to-audio tasks, models like FoleyGen and Smooth-Foley show that incorporating fine-grained temporal and frame-level features can significantly improve synchronization between visual events and generated sound \cite{mei_foleygen_2024,zhang_smooth-foley_2025}.

Despite these advances, several challenges remain. First, while objective metrics (e.g., FAD, IS, KL divergence) indicate strong performance, subjective evaluations often reveal gaps in perceived realism and quality \cite{ghosal_text--audio_2023, mei_foleygen_2024, leonardis_masked_2025}. This discrepancy suggests that current evaluation methods may not fully capture human auditory perception, emphasizing the need for more robust and perceptually grounded metrics. Second, temporal alignment, particularly in complex, multi-event scenarios, remains a persistent challenge, even for state-of-the-art systems \cite{xie_picoaudio_2025,ren_sta-v2a_2025}.

Another key limitation is the trade-off between controllability and creativity. While some models offer precise control over attributes such as timing, pitch, and energy, others prioritize generative diversity at the expense of predictability \cite{barahona-rios_noisebandnet_2024}. As highlighted in the discussion of audio-to-audio systems and creative support tools, users, especially professional sound designers, value systems that enhance creativity while maintaining a degree of control and reliability \cite{kamath_sound_2024}.

Furthermore, accessibility and scalability are important considerations. AI-driven tools have the potential to democratize sound design by reducing the need for large audio libraries and specialized expertise \cite{filimowicz_doing_2022}. However, many high-performing models still require substantial computational resources and large training datasets, which may limit their adoption in smaller studios or independent projects \cite{evans_stable_2025}.

Finally, multimodal systems represent a promising direction for future research. By integrating multiple input types (e.g., text, video, and audio), these models can generate more coherent and context-aware outputs \cite{tang_any--any_2023, ghosh_querymintai_2024, hao_vag_2025}. However, ensuring consistent alignment across modalities remains a complex challenge that requires further investigation.

\begin{table}[]
\caption{Model distribution by category \& subcategory.}
\label{tab:summary}
\begin{tabular}{llcl}
\rowcolor[HTML]{2E5D8C} 
\multicolumn{1}{c}{\cellcolor[HTML]{2E5D8C}{\color[HTML]{FFFFFF} \textbf{Category}}} & \multicolumn{1}{c}{\cellcolor[HTML]{2E5D8C}{\color[HTML]{FFFFFF} \textbf{Subcategory}}} & {\color[HTML]{FFFFFF} \textbf{\# Models}} & \multicolumn{1}{c}{\cellcolor[HTML]{2E5D8C}{\color[HTML]{FFFFFF} \textbf{Dominant Architecture}}} \\
\rowcolor[HTML]{E8F0F8} 
\textbf{Text-to-Audio}                                                               & Latent Diffusion Models                                                                 & \textbf{6}                               & LDM + language encoders                                                                           \\
\rowcolor[HTML]{E8F0F8} 
\textbf{Text-to-Audio}                                                               & Specialized Control                                                                     & \textbf{3}                               & LDM + temporal/attribute control                                                                  \\
\rowcolor[HTML]{E8F0F8} 
\textbf{Text-to-Audio}                                                               & Application-Specific                                                                    & \textbf{2}                               & LDM + domain adaptation                                                                           \\
\rowcolor[HTML]{E8F5F2} 
\textbf{Visual-to-Audio}                                                             & GAN-Based                                                                               & \textbf{2}                               & GAN + visual encoders                                                                             \\
\rowcolor[HTML]{E8F5F2} 
\textbf{Visual-to-Audio}                                                             & Diffusion/Flow-Based                                                                    & \textbf{4}                               & LDM/RFM + alignment modules                                                                       \\
\rowcolor[HTML]{E8F5F2} 
\textbf{Visual-to-Audio}                                                             & Pipeline/Hybrid                                                                         & \textbf{7}                               & Multi-component pipelines                                                                         \\
\rowcolor[HTML]{FDF0E8} 
\textbf{Audio-to-Audio}                                                              & Signal Processing                                                                       & \textbf{1}                               & DDSP-derived filter-bank                                                                          \\
\rowcolor[HTML]{FDF0E8} 
\textbf{Audio-to-Audio}                                                              & Creative Tools                                                                          & \textbf{1}                               & GAN + human-in-the-loop                                                                           \\
\rowcolor[HTML]{F0E8F5} 
\textbf{Multimodal}                                                                  & Cross-Modal / Toolkits                                                                  & \textbf{4}                               & LDM + latent alignment                                                                           
\end{tabular}
\end{table}

\section{Conclusion}
This chapter examined recent developments in AI-based generative models for sound effect synthesis, focusing on text-to-audio, visual-to-audio, audio-to-audio, and multimodal approaches. The findings demonstrate that AI has significantly advanced the capability to generate realistic, diverse, and context-aware sound effects, offering transformative potential for applications such as video games, film production, virtual reality, and interactive systems. These advancements imply a fundamental shift in sound effect synthesis workflows, where generative AI can augment or partially replace traditional manual sound design, enabling rapid prototyping, iterative creativity, and on-demand audio generation tailored to specific scenes or user interactions.

Diffusion-based models and multimodal learning frameworks have emerged as dominant approaches, enabling improved audio quality, semantic relevance, and adaptability. These technologies reduce the reliance on manual sound design processes and large curated libraries, making high-quality audio production more efficient and accessible. As a result, sound effect creation is becoming increasingly democratized, allowing smaller studios and individual creators to produce professional-grade audio content without extensive resources, while also enabling adaptive and personalized soundscapes in real-time applications.

However, challenges remain in achieving perfect temporal synchronization, improving perceptual realism, and balancing user control with generative flexibility. Additionally, the gap between objective evaluation metrics and subjective human perception highlights the need for more comprehensive assessment frameworks. This underscores the continued importance of human oversight, creative direction, and hybrid workflows that combine AI generation with expert refinement to ensure artistic intent and narrative coherence are preserved.

Overall, AI-driven audio generation represents a rapidly evolving field with significant implications for the future of sound design. As these technologies continue to mature, they are likely to play a central role in shaping next-generation digital experiences, enabling richer, more immersive, and more dynamic auditory environments. This evolution points toward a future where sound design becomes more interactive, context-aware, and scalable, fundamentally redefining the role of sound designers from content creators to creative directors of generative systems.

\bibliographystyle{spmpsci}
\bibliography{mybib}
\end{document}